\documentstyle[aps]{revtex}
\input{psfig}

\begin{document}

\textheight 8.8in
\textwidth 6.5in
\topmargin -.25in
\oddsidemargin -.25in
\evensidemargin 0in
\baselineskip 14pt
\def\hm{\ \rm {\it h}^{-1} Mpc}

\title{{\large {\bf Early time perturbations behaviour in scalar 
field cosmologies}}}

\author{Francesca Perrotta$^{1}$ \& Carlo Baccigalupi$^{2}$}
\address{
$^{1}$ SISSA/ISAS, Via Beirut 4, 34014 Trieste, Italy;\\
$^{2}$ INFN and Dipartimento di Fisica, Universit\`a di Ferrara, 
Via del Paradiso 12, 44100 Ferrara, Italy.} 

\baselineskip 10pt
\maketitle
\begin{abstract}  
We consider the problem of the initial conditions and behaviour 
of the perturbations in scalar field cosmology with general 
potential. 

We use the general definition of adiabatic and 
isocurvature conditions to set the appropriate 
initial values for the perturbation in the scalar 
field and in the ordinary matter and radiation components. 
In both the cases of initial adiabaticity and isocurvature, 
we solve the Einstein and fluid equation at early times and 
on superhorizon scales to find the initial behaviour of the 
relevant quantities. 
In particular, in the isocurvature case, we consider models 
in which the initial perturbation arises from the matter 
as well as from the scalar field itself, provided that the initial 
value of the gauge invariant curvature is zero. 

We extend the standard code to include all these cases, 
and we show some results concerning the power spectrum of 
the cosmic microwave background temperature 
and polarization anisotropies. In particular, it turns out that 
the acoustic peaks follow opposite behaviours in the adiabatic 
and isocurvature regimes: in the first case 
their amplitude is higher than in the corresponding 
pure cold dark matter model, while they make the opposite thing 
for pure isocurvature initial perturbations. 
\end{abstract}

\section{Introduction}
\label{introduction}

In recent times, the need for a "quintessence" component has come out due
to the several difficulties  of the standard $\Omega_m=1$ Cold Dark Matter
(CDM) model which proved unable to  explain  the observed features of
large scale structure.
In the context of inflationary cosmologies, we expect that the present
spatial curvature of the Universe is negligible and the total energy
density equals the critical energy density; on the other hand,
there is growing observational evidence that the matter energy density is
remarkably below the critical value, even taking into account exotic and
so far undetectable particles known as cold dark matter.  
Thus, we are faced with figuring out how to explain missing-energy values
of as much as 70 or 80 percent of critical density. 

Further, there is need
to have the age of the Universe, $t_0$,  exceeding the age of globular
clusters in our galaxy; the limits on $t_0$ are holding at about 13 
Gyr or more \cite{AGE}, 
and when combined with current estimates of the Hubble expansion
parameter, converging from different methods to $H_0 \approx 60 \pm 10 
$km/sec/Mpc \cite{HUBBLE}, 
give rise to an observed value of the "expansion-age"
parameter $H_0 t_0\simeq .8$, sensibly 
higher than 2/3 as predicted by the standard Einstein-De
Sitter model. 

Preserving the flatness of the Universe, its age 
could be enhanced by lowering the matter content in models involving a
component whose equation of state is different from matter and radiation,
for example in models including a cosmological constant. 
One more problem with the CDM model arises from the
mismatching of the galaxy clustering power spectrum shape, when only COBE  
normalized spectra are considered. All this motivations have ratified 
the demise of the standard CDM model, leaving cosmologists with the open
question of what the missing-energy candidate could be. 

Apart from the cosmological constant,  introduced in some models in order
to maintaining the spatial flatness,  but still retaining serious unsolved
theoretical issues, several "quintessence" models where  
proposed as candidates for the missing energy, often modeled as scalar
fields rolling down their potentials
\cite{VL,FJ,CLW,CDF,AC,ZWS,HWDCS,RQ,RP}, 
or more generally described in terms of an unspecified equation of 
state different from that of matter and radiation 
\cite{H,CDS,TW}.  We refer to any component whose properties are 
well described in terms of a scalar field evolving in a potential which
couples to ordinary matter only through gravity. In some sense, it can
behave like a cosmological constant when its kinetical energy is
negligible with respect to the potential energy, so that the scalar field
equation of state approaches -1; due to the strongly relativistic nature
of such component, the characteristic scale of clustering processes for a
scalar field is just the horizon, \cite{RP} giving a
similarity with a cosmological constant in the undetectability of
quintessence energy concentrations on scales smaller than the horizon.   

The interesting  feature of the "quintessence"
component  is  just that, contrary to the cosmological constant,  it is
time-varying and spatially inhomogeneous, so that it can develop
fluctuations which can be relevant in the 
perturbation growth and can leave a characteristic signature in the 
CMB and in the large scale structure. Even though  many of these imprints
have been caught in previous works, the issue of initial conditions 
and scalar field perturbations has often been underestimated; 
in particular, we found a gap regarding the opportunity to impose 
isocurvature initial conditions in a several-components system 
including a minimally coupled scalar field.
Our aim is to give a complete prescription for describing adiabatic and
isocurvature initial conditions if an additional component is present in
the form of such scalar field; this can be acquired by giving the set of
equations relating all the fluid components needed in the two cases.

In order to do that, we need the background and perturbation equations 
which we briefly review on section II and section III. 
The original results of our work are presented in sections IV and V,
where we work in the formalism of the synchronous gauge, by generalizing 
the work of Ma and Bertschinger \cite{MB} and finding the
super-horizon-scale behaviour of perturbations at early times, starting
from initial zero entropy perturbations (adiabatic case) or initial zero
curvature perturbations (isocurvature case). We express the needed 
gauge-invariant quantities, namely entropy and curvature
perturbations, in terms of synchronous perturbations of baryons, 
photons, massless neutrinos, cold dark matter
and a minimally-coupled scalar field, as well as metric perturbations.  
In section VI the results are translated in conformal Newtonian gauge, 
and in section VII  we numerically investigate the growth of entropy 
and curvature perturbations starting from different initial conditions, 
and we compare them with the corresponding behaviours in the standard CDM 
model. In that computation, the adopted scalar field is associated with 
an ultralight pseudo Nambu-Goldstone boson 
\cite{nambu}, with global spontaneous symmetry 
breaking scale $f \simeq 10^{18}$ GeV
and explicit breaking scale $M \sim 10^{-3} $ eV; such a field 
should be acting at present like  an effective cosmological
constant and  dominating the energy density of the Universe. 
Also we plot, and discuss, the pure adiabatic and pure
isocurvature CMB anisotropies spectra, again making a comparison with
the standard CDM.

\section{Einstein and conservation equations}

We begin by a brief review of an homogeneous Friedmann Robertson 
Walker (FRW) cosmology in which there is the additional contribution coming 
from a minimally coupled real scalar
field $\phi$ evolving in a potential $V(\phi)$.

We consider only models with total $\Omega =1$ in this paper , and we work
in conformal coordinates, so that the line element is 
$ds^2=a^2 [-d\tau ^2 + \delta_{i j }dx^i dx^j]$ where $a$ is the cosmic
scale factor and $\tau$ is the conformal time.\\
The scalar field  energy density and pressure, associated to the
Lagrangian describing the classical behavior of $\phi$
\begin{equation}
{\cal{L}} = - {1 \over 2} \sqrt{-g} \left[ 
g^{ \mu \nu }  \partial_{\mu} \phi \partial_{\nu} \phi  + 2V(\phi)
\right]\ ,
\end{equation}
follow from expression of the stress-energy tensor
\begin{equation}
T ^{\mu}_{\nu}= \phi ^{;\mu}\phi_{;\nu}-{1 \over 2}
(\phi ^{;\alpha}\phi_{;\alpha}+2V)\delta^{\mu}_{\nu}\ ,
\end{equation}
and are given by
\begin{equation}
\label{rhopphi}
 \rho_{\phi}={ 1 \over 2a^{2}} \dot{\phi}^2 + V(\phi) \  ; \ \ \
  p_{\phi}={ 1 \over 2a^{2}} \dot{\phi}^2 - V(\phi)   \ ,
\end{equation}
where the overdot denotes the derivative with respect to conformal time
$\tau$. The above quantities evolve according to the Friedmann equation , in which we separate 
the contributes  of matter, radiation and scalar field to the total energy density,
\begin{equation}
\label{Friedmann}
{\cal H }^2=\left({\dot{a}\over a}\right)^{2}={8\pi G\over
3}a^{2}\left[
\rho_{m}+\rho_{r}+\rho_{\phi} 
\right]\ ,
\end{equation}
together with the conservation equations 
\begin{equation} 
\label{KG}
\ddot{\phi}+2 {\cal{H}} \dot{\phi}+a^{2}{dV\over d\phi}=
{1 \over  a^2} {d \over d \tau }(a^2 \dot{\phi} ) + a^2 V'(\phi)
=0\ ,
\end{equation}
\begin{equation}
\label{conservation}
\dot{\rho}_{n}+n {\cal{H}} {\rho}_{n} = 0\ ,
\end{equation}
where $\cal{H}$ is the conformal expansion rate of the Universe, 
${\rho}_{n}$ is the energy density contributed by radiation $(n=4)$ or
nonrelativistic matter $(n=3)$ and $V'=dV/d\phi$. Note
that, from (\ref{rhopphi}),  the second order Klein-Gordon
equation (\ref{KG}) is equivalent to the conservation law 
\begin{equation}
\dot{\rho}_{\phi}=-3 {\cal H}({\rho}_{\phi}+p_{\phi})  \ \ .
\end{equation} 
Including all the modifications due to the additional
scalar field component  , we shall carry out a fully relativistic
treatment of the perturbations of this background , based on the
notation of Ma and  Bertschinger \cite{MB}.
We work in the Fourier
space and we perform the parametrization  of the perturbed
quantities in the formalism of the synchronous gauge , in which
the perturbed line element is
$ds^2=a^2[-d \tau  ^2 +( \delta _{i j} + h  _{i j})dx^i dx^j] $.
Since we are interested here only on scalar-type perturbations,
the metric perturbations can be parametrized as
\begin{equation}
h_{i j}( {\bf x}, \tau ) = \int d^3 k e^{i {\bf k} \cdot {\bf x}}
\left[ 
{\bf {\hat{k}_i  \hat{k}_j}}  h( {\bf k}, \tau)
+ ( {\bf {\hat{k}_i  \hat{k}_j}}- {1 \over 3} \delta_{ i j} )
6 \eta  ( {\bf k}, \tau)\right]\ ,
\end{equation}
with $ {\bf k}=k \hat{k} $ and $h$ denoting the trace of $h_{ij}$.
\\
Note that the synchronous potentials $h$ and $\eta$ in $k$-space are
related to the gauge-invariant variables ${\bf \Phi}_A $ of
Bardeen $(1980)$ and ${\bf \Psi}$ of Kodama and Sasaky $(1984)$ by
the
relation
\begin{equation}
\label{psiA}
{\bf \Phi}_A = {\bf \Psi} ={1\over
2k^{2}}\left[\ddot{h}+6\ddot{\eta}+
{\dot{a}\over a}\left(\dot{h}+6\dot{\eta}\right)\right]\ ,
\end{equation}
which allows us to relate $h$ and $\eta$ to the gauge-invariant
curvature perturbation $\zeta$ (see \cite{MFB})
\begin{equation}
\label{gicurvature}
\zeta = {2 \over 3}({\cal H}^{-1} \dot{\bf \Psi} + {\bf \Psi}) / (1+w) +
{\bf \Psi}\ ,
\end{equation}
being $w=p/\rho$; the above expression
will
be useful in the following. Now, focus on the equations describing
the evolution of perturbations involving the various components.


As it is well known, the scalar field can mimic a cosmological constant 
if its kinetic energy is negligible with respect to the potential one. 
However, a substantial difference is that it admits perturbations around 
the homogeneous solution of (\ref{KG}); in linear theory, 
they are described by small fluctuations $\delta\phi$ and 
$\dot{\delta\phi}$ around the background values, driven by the equation 
of motion
\begin{equation}
\label{KGp}
\ddot{\delta\phi}+2{\dot{a}\over a}\dot{\delta\phi}-
\nabla^{2}\delta\phi +a^{2}{d^{2}V\over d\phi^{2}}\delta\phi+
{1\over 2}\dot{\phi}\dot{h}=0\ . 
\end{equation} 
The density, pressure and velocity perturbations for the scalar field are 
described as usual by the following quantities
\begin{eqnarray}
\label{drhophidpphi}
\delta\rho_{\phi}=-\delta T_{0}^{0} &=&{\dot{\phi}\delta\dot{\phi}
\over a^{2}}+V'\delta\phi\ \ , \\ 
\delta p_{\phi}={1\over 3}\delta T_{i}^{i} &=& {\dot{\phi}\delta\dot{\phi} 
\over a^{2}}-V'\delta\phi\ \ , \\
({\rho}_{\phi}+p_{\phi}){\theta}_{\phi}=\delta T_i^0 &=& k a^{-2} \dot{\phi} 
\delta \phi\ \ , \\
p_{\phi} {\pi}_{\phi} &=& 0\ \ , 
\end{eqnarray}
\begin{equation}
\label{deltaphithetaphi}
\delta_{\phi}={\delta\rho_{\phi}\over\rho_{\phi}}\ \ ,\ \ 
\theta_{\phi}={k\over a^{2}}
{\dot{\phi}\delta\dot{\phi}\over\rho_{\phi}+p_{\phi}}=
k{\delta\dot{\phi}\over\dot{\phi}}\ ;
\end{equation}
therefore, we define the differential ratio 
\begin{equation}
\label{dpdrhophi}
{\delta p_{\phi}\over \delta\rho_{\phi}}={\dot{\phi}\delta\dot{\phi}-
a^{2}V'\delta\phi\over \dot{\phi}\delta\dot{\phi}+
a^{2}V'\delta\phi}=c_{\phi}^{2}+{p_{\phi}\over\delta\rho_{\phi}}
\Gamma_{\phi}\ ,
\end{equation}
that differs from the scalar field sound velocity 
\begin{equation}
\label{cphi}
c_{\phi}^{2}={\dot{p}_{\phi}\over\dot{\rho}_{\phi}}
\end{equation}
by the term $\Gamma_{\phi}$ describing  
the entropy contribution \cite{KS}.


It is useful to describe radiation in terms of the coefficients characterizing 
the Legendre expansion of the temperature  and polarization brightness  functions, 
$\Delta_{T}( \bf{k}, {\hat{\bf{n}}}, \tau)$ and $\Delta_{P}( \bf{k}, 
{\hat{\bf{n}}}, \tau)$ :

\begin{equation}
\label{DeltaT}
\Delta_{T}( {\bf{k}, {\hat{\bf{n}}}}, \tau)= {\sum_{ {l}=0}^{\infty}}
 (- { {i}}^{ { l}}) 
(2 {l} + 1 ) 
\Delta_{T  l}( k, \tau) P_{ l} ({\hat{\bf{k}}} \cdot {\hat{\bf{n}}} )\ ,
\end{equation}
\begin{equation}
\Delta_{P}( {\bf{k}, {\hat{\bf{n}}}}, \tau)= {\sum_{ {\cal l}=0}^{\infty}}
 (- { i}^{ {l}}) 
(2 {l} + 1 ) 
\Delta_{P l}( k, \tau) P_{l} ({\hat{\bf{k}}} \cdot {\hat{\bf{n}}})\ .
\end{equation}
Their evolution is completely determined by the Boltzmann equations; 
denoting by ${\sigma}_T $ the Thomson scattering cross section and by 
$n_e$ the electron density, we have for photons:

\begin{eqnarray}
\label{deltag}
{\dot{\delta}}_{\gamma}&=&-{4\over 3}\theta_{\gamma}-{2 \over 3}\dot{h}\ , \\
\label{thetag}
\dot{\theta}_{\gamma} &=& k^2 \left( {1 \over 4} {\delta}_{\gamma} - 
{\sigma}_{\gamma} \right )
+a  n_e {\sigma}_T  ( {\theta}_b - {\theta}_{\gamma} )\ , \\
\label{sigmag}
2 {\dot {\sigma} }_{\gamma}  &=& {8 \over 15} {\theta}_{\gamma} - {3 \over 5 } k {\Delta}_{\gamma 3 }
+{4 \over 15} \dot{h} + {8 \over 5 } \dot{\eta}   
-{9 \over 5 } a n_e {\sigma}_T {\sigma}_{\gamma} + 
{1 \over 10}a n_e {\sigma}_T 
({\Delta}_{P0 (\gamma)} + {\Delta}_{P 2 (\gamma)} )\ ,\\
\label{deltatl}
\dot{\Delta}_{T { l}(\gamma)} &=& { k \over 2 { l} +1 } 
[ {l }{\Delta}_{T ({l} -1)(\gamma)}
-( { l} +1) {\Delta}_{T ({ l} +1)(\gamma)}] - 
a n_e {\sigma}_{T}{\Delta}_{T {l}(\gamma)} 
 \ \ \  ({ l} \ge{3})\ \ , \\
\label{deltapl}
\dot{\Delta}_{P {l}(\gamma)} &=& { k \over 2 { l} +1 } 
[ { l }{\Delta}_{P ({l} -1)(\gamma)}
-( { l} +1) {\Delta}_{P ({ l} +1)(\gamma)}] +  
a n_e {\sigma}_T \left[{1 \over 2} ( {\Delta}_{T 2(\gamma)}  + 
{\Delta}_{P 0(\gamma) } + {\Delta}_{P 2 (\gamma)}) 
\left( {\delta}_{l 0} +  {{\delta}_{{l} 2} \over 5} \right)
-{\Delta}_{P { l}(\gamma)} \right]\ ,
\end{eqnarray}
where 
\begin{equation}
{\delta}_{\gamma}={\Delta}_{ T0} \ ,\ \ 
{\theta}_{\gamma}={3 \over 4} k {\Delta}_{T1} \ ,\ \
{\sigma}_{\gamma}={1 \over 2} {\Delta}_{ T2}\ .
\end{equation} 
The perturbed stress-energy tensor for radiation contributes with the following non-zero quantities:
\begin{eqnarray}
\delta T ^{0}_0 &=& - {\rho}_{\gamma} {\delta}_{\gamma}\ ,\\
ik^i \delta T ^{0}_i &=& {4 \over 3 } {\rho}_{\gamma} {\theta}_{\gamma}\ , \\
\delta T^{i}_{j}&=& {1 \over 3} {\rho}_{\gamma} {\delta}_{\gamma} + 
{\Sigma}^{i}_{j}\ ,\\ 
({\hat{\bf{k}}}_i {\hat{\bf{k}}}_{j}-{1 \over 3} {\delta}_{i j}) 
{\Sigma}^{i}_{j} &=& -{4 \over 3} {\rho}_{\gamma} {\sigma}_{\gamma}\ .
\end{eqnarray}
The expansion (\ref{DeltaT}) also applies for massless neutrinos; 
their  evolution equations in the synchronous gauge  are given by the 
following system:
\begin{eqnarray}
\label{deltanu}
\dot{\delta}_{\nu}&=&-{4 \over 3} {\theta}_{\nu} -{2 \over 3 } \dot{h}\ , \\
\label{thetanu}
\dot{\theta}_{\nu}&=&k^2 \left( {1 \over 4} {\delta}_{\nu} -{\sigma}_{\nu} 
\right)\ ,\\
\label{sigmanu}
\dot{\Delta}_{T 2 (\nu)}&=&2 \dot{\sigma}_{\nu} =
{8 \over 15} {\theta}_{\nu} -{3 \over 5} k {\Delta}_{T3(\nu)} +
{4 \over 15} \dot{h} + {8 \over 5 } \dot{\eta}\ , \\
\label{deltatlnu}
\dot{\Delta}_{T l (\nu)}&=& {k \over 2l+1} [l {\Delta}_{T (l-1) (\nu)}- (l+1) 
{\Delta}_{T (l+1) (\nu)} ]   \, \ \  \ (l \ge{3})\ .
\end{eqnarray}
Pressureless Cold Dark Matter interacts only gravitationally with other 
particles and in the synchronous gauge its peculiar velocity is zero;  
setting ${\theta}_c=0$, the evolution of CDM density perturbations is 
given by
\begin{equation}
\label{deltacdot}
\dot{\delta}_c = -{1 \over 2 } \dot{h}\ ,
\end{equation}
and the non-zero component of its perturbed stress-energy tensor is
\begin{equation}
\label{deltac}
\delta T_0^0= -{\rho}_c {\delta}_c\ .
\end{equation}
Taking into account the coupling between photons and 
baryons by Thomson scattering, 
\begin{eqnarray}
\label{deltab}
\dot{\delta}_b &=& - {\theta}_b -{1 \over 2} \dot{h}\ ,\\
\label{thetab}
\dot{\theta}_b &=& { \dot{a} \over a } {\theta}_b+c_s^2 k^2{\delta}_b +
{4 {\rho}_{\gamma}  \over 3 {\rho}_b } 
a n_e {\sigma}_T ( {\theta}_{\gamma}-{\theta}_b )\ ,
\end{eqnarray}
and the perturbed stress-energy tensor for baryons contributes by
\begin{eqnarray}
\delta T_0^0 &=& -{\rho}_b {\delta}_b\ ,\\
i k^i T_i^0 &=& {4 \over 3 } {\rho}_{\gamma} {\theta}_b\ .
\end{eqnarray}
All these ingredients are to be implemented in perturbed Einstein equations
\begin{eqnarray}
\label{t00}
k^{2}\eta -{1\over 2}{\cal H}\dot{h} &=& 4\pi Ga^{2}\delta T_{0}^{0}\ ,\\
\label{ti0}
k^{2}\dot{\eta} &=& 4\pi G a^{2}ik^{i}\delta T_{i}^{0}\ ,\\
\label{tii}
\ddot{h}+2{\cal H}\dot{h}-2k^{2}\eta &=& -8\pi G a^{2} \delta T_{i}^{i}\ ,\\
\label{tij}
\ddot{h}+6\ddot{\eta}+2{\cal H}(\dot{h}+6\dot{\eta})-2k^{2}\eta &=&
24\pi Ga^{2}\left(\hat{\bf k}_{i}\hat{\bf k}_{j}-{1\over 3}
\delta_{ij}\right)\Sigma_{j}^{i}\ .
\end{eqnarray}
This system of differential equations can be integrated once fixed
the appropriate initial conditions, which  will be the content of the
next sections.

\section{Initial conditions and superhorizon evolution}

In order to start the numerical integration of the evolution equations
wrote down in the previous section, one has to impose appropriate 
initial conditions to the fluid and metric perturbations.
Although a general perturbation need not to be either isothermal
(entropic) or adiabatic (isoentropic), 
it can always be expressed as a linear superposition of 
adiabatic and isothermal components \cite{PAD}. 
Also, it is useful to recall that isocurvature perturbations 
may be present in this kind of models \cite{M}. 
We explore both these conditions in scalar field cosmology. In particular 
we search for the initial values of the field perturbations 
$ \delta\phi_{0}$ and $\delta\phi_{t 0}$ (initial real time derivative) 
that realize initial adiabaticity and isocurvature, together with  
appropriate initial conditions on the other perturbed quantities.
For this purpose, we will focus on the 
initial fluctuation of the real time derivative 
of the scalar field perturbation, since the conformal time 
derivative is always zero at $a=0$ by definition 
($\delta\dot{\phi}=a\cdot\delta\phi_{t}$). 
The same happens of course to the initial conformal time 
derivative of the background scalar field  $\phi$; 
if in general the latter has 
an initial non-vanishing kinetic energy, so that $\phi_{t 0}=
(d\phi /dt)_{0}\ne 0$, its conformal time velocity 
$\dot{\phi}_{0}$ is zero since $\dot{\phi}=a\cdot\phi_{t}$.

In the following, we will need to use    
the scale factor behaviour at early times, when $a\ll 1$. 
We will often use the expansion of the scale factor in powers of the
dimensionless parameter $\epsilon$:
\begin{equation}
\label{epsilon}
\epsilon =\sqrt{8\pi G\rho_{c}\Omega_{r}\over 3}\tau =
H_{0}\sqrt{\Omega_{r}}\tau ={\cal C}\tau\ ,
\end{equation}
where $\rho_{c}$ and $H_{0}$ are the present critical density and 
Hubble parameter respectively, and 
$\Omega_{r}=\Omega_{\gamma}+\Omega_{\nu}$ is the total
radiation density contribution at the present; indeed, as it can 
be easily verified, being  the scale factor 
$a\ll 1$ at early times, 
we can neglect the scalar field contribution in 
equation (\ref{Friedmann}), that admits the simple solution
\begin{equation}
\label{aepsilon}
a(\epsilon )= \epsilon +{1\over 4}
{\Omega_{m}\over\Omega_{r}}\epsilon^{2}+O(\epsilon^{3})\ ; 
\end{equation}
besides, the expansion rate behaves  as
\begin{equation}
\label{Hepsilon}
{\cal H}={\cal{C}} \left[{1\over\epsilon}+
{1\over 4}{\Omega_{m}\over\Omega_{r}}-
{1\over 16}\left({\Omega_{m}\over\Omega_{r}}\right)^{2}
\epsilon +O(\epsilon^{2})\right]\ .
\end{equation}
Before to go on, it is worth to remind some general results concerning
the synchronous gauge behaviour of metric and density
perturbations on superhorizon scales (we refer to the work of
Ma and Bertschinger \cite{MB}, although they did not include the scalar
field component ).
We impose initial condition at an  early time, deep
in the radiation era, when photons and baryons are tightly coupled and
can be considered as a single coupled fluid; due to the large Thomson
scattering opacity, the $l \ge{2} $ moments of the photon temperature
brightness
function (\ref{deltatl}) (in particular, the shear $\sigma_{\gamma}$)  and
the polarization brightness function
(\ref{deltapl})  are driven to zero; similarly, to the lowest order in
$k \tau$, one can ignore the $l \ge{3} $  moments of the neutrino 
temperature brightness function .
Thus, the equations
(\ref{deltag}),(\ref{thetag}), (\ref{deltanu}),(\ref{thetanu}) become:
\begin{eqnarray}
\label{system1}
\dot{\delta}_{\gamma}+{4 \over 3} {\theta}_{\gamma}+{2 \over 3} \dot{h}=0
\ \ &,&\ \ 
\dot{\theta}_{\gamma}-{1 \over 4}k^2 {\delta}_{\gamma}=0\ ,\\
\label{system2}
\dot{\delta}_{\nu}+{4 \over 3} {\theta}_{\nu}+{2 \over 3} \dot{h}=0 
\ \ &,&\ \ \dot{\theta}_{\nu}-{1 \over 4}k^2 
({\delta}_{\nu}-4  {\sigma}_{\nu})=0\ ,\\
\label{system3}
\dot{\sigma}_{\gamma}-{2 \over 15} (2{\theta}_{\gamma}+
\dot{h} +6 \dot{\eta})=0\ \ &,&\ \ 
\dot{\sigma}_{\nu}-{2 \over 15} (2{\theta}_{\nu}+\dot{h} 
+6\dot{\eta})=0\ .
\end{eqnarray}
When we impose initial conditions, at $\epsilon\ll 1 $, to get starting
values for numerical integration,  all the $k-$ modes are still outside
the horizon, i.e. $k\ll a{ H}=1/\tau$ (the last equality 
holds in a radiation dominated universe).
Our aim is to extract the analytical
time-dependence of superhorizon-sized perturbations at early times, once
the initial conditions are realized: thus we find the early 
time form of equations 
(\ref{t00})-(\ref{tij}), (\ref{system1}-\ref{system3}), (\ref{KGp}) and 
we find their solutions in successive powers of $k\tau$. 
To set-up the growth of perturbations, we must assume that 
at least a single perturbation is nonzero  at initial time , in order to
generate  all the others. 

\section{Adiabatic initial conditions}

The first necessary step to impose adiabatic conditions is setting to zero
the initial entropy perturbation; ultimately, the origin of this result is
that there is initially a single curvature perturbation (generated we
suppose by inflation) and all later perturbations are inherited from it. 
The entropy exchange between
any two fluid species $a$ and $b$ is ruled by the gauge invariant quantity
\begin{equation}
\label{sab}
S_{ab}={\delta_{a}\over 1+w_{a}}-{\delta_{b}\over 1+w_{b}}\ ,
\end{equation}
that must be set to zero initially \cite{KS}.
The second request comes from setting to zero the first
time derivative of $S_{ab}$; actually,
$S_{ab}$ obeys the following differential equation:
\begin{equation}
\label{entropy}
\dot{S}_{ab}=-kV_{ab}-3{\cal H}\Gamma_{ab}\ ,
\end{equation}
where $\Gamma_{ab}$ is defined as
\begin{equation}
\label{Gammaab}
\Gamma_{ab}={w_{a}\over 1+w_{a}}\Gamma_{a}-
{w_{b}\over 1+w_{b}}\Gamma_{b}\ ,
\end{equation}
$\Gamma_{a}$ being the gauge-invariant amplitude of the entropy
perturbation of the fluid species $a$. The 
quantity $V_{ab}=v_{a}-v_{b}$ is the gauge-invariant difference between
the gauge-dependent velocity perturbations of the species $a$ and $b$.
In order to have adiabatic initial conditions, both these terms on the
right hand side
of equation (\ref{entropy}) are initially set to zero. 
Thus, for each pair of fluid components, we impose 
\begin{equation}
\label{adiab}
S_{ab}=\dot{S}_{ab}=0\ .
\end{equation}
In particular, applying (\ref{adiab}) to the  scalar field and
another
component (that we leave unspecified and label  
with $x$),  will relate the initial values of $\delta \phi$ and $\delta
\dot {\phi} $ to  the other energy component and metric perturbations.   
 The first condition in 
(\ref{adiab}) gives:
\begin{equation}
\label{sab=0}
\phi_{t}\delta\phi_{t}+V'\delta\phi =
\phi_{t}^{2}{\delta_{x}\over 1+w_{x}}\ .
\end{equation}
Posing $\dot{S}_{\phi x}=a\cdot S_{\phi x t}=0$
and using the Klein Gordon equation (\ref{KG},\ref{KGp}),
we obtain 
\begin{equation}
\label{spab=0}
\delta\phi_{t}\left(1-{k^{2}\phi_{t}\over 6a^{2}HV'}\right)={1\over 6H}\left[
-{1\over 2}\phi_{t}h_{t}-\left({\delta_{x}\over 1+w_{x}}\right)_{t}
\phi_{t}+{\delta_{x}\over 1+w_{x}}\left(6H\phi_{t}+2V'-
{k^{2}\phi_{t}^{2}\over a^{2}V'}\right)\right]\ .
\end{equation}
Combining them together we find:
\begin{equation}
\label{deltaphiad}
\delta\phi ={1\over V'}
\left(\phi_{t}^{2}{\delta_{x}\over 1+w_{x}}-
\phi_{t}\delta\phi_{t}\right)\ ,
\end{equation}

\begin{equation}
\label{deltaphidotad}
\delta\phi_{t}={1\over 6H-  k^{2}\phi_{t}/(a^{2}V') }\left[
-{1\over 2}\phi_{t}h_{t}-\left({\delta_{x}\over 1+w_{x}}\right)_{t}
\phi_{t}+{\delta_{x}\over 1+w_{x}}\left(6H\phi_{t}+2V'-
{k^{2}\phi_{t}^{2}\over a^{2}V'}\right)\right]\ .
\end{equation}
The above expressions specify the general adiabatic
conditions for the scalar field. 
Now,  let us make a link to previous
works; in \cite{MB} the adiabatic initial values and early time
behaviours of the matter and the radiation components were found in
the synchronous gauge; these results apply here, too.
Indeed, as it can be easily seen from the
Einstein equations, the contribution of the scalar field fluctuations
is negligible at early times $a\ll 1$ with respect to the matter
and radiation ones, by a factor $a^{3}$ and $a^{4}$ respectively.
Thus the approximations and treatment developed in \cite{MB}
is valid also here for what concerns the ordinary fluid components,
i.e. photons, massless neutrinos, baryons and dark matter; the
time-dependence of   the  resulting superhorizon-sized perturbations 
$(k \tau \ll 1 )$   is found by expanding the Einstein
equations into powers of $k\tau$ and resolving the system of coupled
differential equations to obtain  the leading-order terms : 
\begin{equation}
\label{mbdelta}
\delta_{\gamma}=\delta_{\nu}={4\over 3}\delta_{b}={4\over 3}
\delta_{c}=-{2\over 3}{\cal N}(k\tau )^{2}\ ,
\end{equation}
\begin{equation}
\label{mbtheta}
\theta_{\gamma}={15+4R_{\nu}\over 23+4R_{\nu}} \ \theta_{\nu}=
\theta_{b}=-{1\over 18}{\cal N}k^{4}\tau^{3}\ ,\ \ \theta_c=0 \ \ , \
\end{equation}
\begin{equation}
\sigma_{\gamma}=0\ ,\ \sigma_{\nu}={4{\cal N}\over 45+12R_{\nu}}
(k\tau )^{2}\ ,
\end{equation}
\begin{equation}
\label{mbheta}
h={\cal N}(k\tau )^{2}\ , \ \eta =2{\cal N}-
{5+4R_{\nu}\over 90+24R_{\nu}}{\cal N}(k\tau )^{2}\ ,
\end{equation}
where $R_{\nu}=\rho_{\nu}/(\rho_{\nu}+\rho_{\gamma})$ and 
${\cal N}$ is a normalization constant. 
Using these results, it is immediate to see from (\ref{deltaphiad},
\ref{deltaphidotad})
that, imposing adiabatic initial  conditions, the initial values of
$\delta \phi$ and $\delta \phi_t $ must be set to zero. Adiabatic
conditions can be strictly  verified $\it{only}$ at this initial time,
due to the effect of the mutual coupling between total  density
perturbations and entropy perturbations which appear in a generic
multi-component fluid. Starting from initial zero values and 
using (\ref{mbheta}), $\delta \phi$
and  $\delta \dot{\phi}$ will evolve according to the (\ref{KGp}) which
can be easily integrated once dropped terms of highest order in $\tau$;
this gives the following  behaviours at early times 
($ a \ll  1$):
\begin{equation}
\label{adiniphi}
\delta {\phi}=-{1\over 20}\phi_{t 0}\ {\cal N}{\cal C}k^{2}\tau^{4}\ ,
\end{equation}
\begin{equation}
\label{adiniphidot}
\delta \dot{\phi}=-{1\over 5}{\phi}_{t 0}\ {\cal N}{\cal C}k^2 \tau^{3}\ ,
\end{equation}
having considered the lowest terms in $\tau$, thereby 
approximating  the time derivative of $\phi$ with its
value at the initial time ${\phi}_{t 0}$  .
We have inserted these inputs into the standard CMB code and
in Section VII we shall  expose some numerical results. 
Now, let us turn to the second class of initial conditions. 

\section{Isocurvature initial conditions}

The isocurvature initial conditions are obtained by
setting to zero the gauge invariant 
curvature perturbation. Its expression is given in terms of
the gauge invariant perturbation potential $\Psi$ \cite{MFB}:
\begin{equation}
\label{curv}
\zeta ={2\over 3}
\left({{\cal H}^{-1}\dot{\Psi}+\Psi\over 1+w}\right)+\Psi\ ;
\end{equation}
 we point out  that here, as in \cite{KS,MB},
$\Psi$ indicates  
the gauge invariant $\Phi_{A}$ of the original work of Bardeen
\cite{Bardeen}, while in \cite{MFB} the same quantity it is indicated as
$\Phi$.
Its expression in terms of the metric perturbations $h$ and $\eta$
in the synchronous gauge is:
\begin{equation}
\label{psisync}
\Psi ={1\over 2k^{2}}\left[\ddot{h}+6\ddot{\eta}+
{\dot{a}\over a}\left(\dot{h}+6\dot{\eta}\right)\right]\ . 
\end{equation}
Therefore, the appropriate isocurvature initial
conditions are realized by the time growing solutions of
the system (\ref{t00})-(\ref{tij}) in which $\Psi$ and 
${\cal H}^{-1}\dot{\Psi}$ are zero initially. 
First, let us see that, 
if the variables describing all the perturbations 
are regular enough to be derivable at least four times 
in $\tau =0$, then the isocurvature initial conditions are 
simply imposed by setting the metric and radiation perturbations 
to zero initially. 
\begin{equation}
\label{iso}
{\rm Isocurvature   \ \ initial  \ \ conditions:}\
h_{0}=\eta_{0}=
\delta_{\gamma}=\theta_{\gamma}=\sigma_{\gamma}=
\delta_{\nu}=\theta_{\nu}=\sigma_{\nu}=0\ .
\end{equation}
This can be easily seen by using essentially the 
Einstein equation (\ref{tij}); multiplying both the members by $a^{4}$, 
deriving once and factoring out the present critical density 
$\rho_{c}$, it takes the following form: 
\begin{equation}
\label{a4e4}
a^{2}{d^{3}\over d\tau^{3}}\left(h+6\eta\right)+
4a\dot{a}\left(\ddot{h}+6\ddot{\eta}\right)+
\left(2\dot{a}^{2}+2a\ddot{a}\right)
\left(\dot{h}+6\dot{\eta}\right)
-4k^{2}a\dot{a}\eta -2k^{2}a^{2}\dot{\eta}=
-32\pi G\rho_{c}\left(\Omega_{\gamma}\dot{\sigma}_{\gamma}+
\Omega_{\nu}\dot{\sigma}_{\nu}\right)\ .
\end{equation}
Since by hypothesis $h$ and $\eta$ are derivable four times in 
$\tau =0$, $h+6\eta$ admits the following early time expansion 
\begin{equation}
\label{hplus6}
(h+6\eta)(\tau )={d\over d\tau}(h+6\eta)_{0}\tau +
{1\over 2}{d^{2}\over d\tau^{2}}(h+6\eta)_{0}\tau^{2} +
{1\over 6}{d^{3}\over d\tau^{3}}(h+6\eta)_{0}\tau^{3} +
{1\over 24}{d^{4}\over d\tau^{4}}(h+6\eta)_{0}\tau^{4} +
O(\tau^{5})\ ,
\end{equation}
since with the initial condition (\ref{iso}) its initial value is zero. 
At $\tau =0$ the only term that survives in (\ref{a4e4}) is 
$\dot{a}^{2}(\dot{h}+6\dot{\eta})$ since 
$\dot{a}^{2}_{0}= 8\pi G\rho_{c}(\Omega_{\gamma}+\Omega_{\nu})/3$. 
Then, by using equations (\ref{system1},\ref{system2},\ref{system3}) 
one obtains:
\begin{equation}
\label{isodim1}
\left(\dot{h}+6\dot{\eta}\right)_{0}{48\pi G\over 5}\rho_{c}
\left(\Omega_{\gamma}+\Omega_{\nu}\right)=0\Rightarrow 
\left(\dot{h}+6\dot{\eta}\right)_{0}=0\ .
\end{equation}
In the same way, by deriving again (\ref{a4e4}) 
one gains 
\begin{equation}
\label{isodim2}
\left(\ddot{h}+6\ddot{\eta}\right)_{0}=0\ .
\end{equation}
Instead, it may be easily see that $d^{3}(h+6\eta )/d\tau^{3}$ 
can be different from zero; in fact, deriving (\ref{a4e4}) 
three times would bring 
$d^{3}(h+6\eta )/d\tau^{3}\propto k^{2}\dot{a}^{2}\dot{\eta}$ 
that may be different from zero by hypothesis (\ref{iso}). 
This means that, for $\tau\rightarrow 0$, 
$h+6\eta = O(\tau^{3})$; since ${\cal H}=1/\tau$ to the 
lowest order, this is evidently enough to make $\Psi_{0}=
({\cal H}^{-1}\dot{\Psi})_{0}=\zeta_{0}=0$, 
showing that the initial conditions (\ref{iso}) imply 
isocurvature. 

It is evident that the initial condition (\ref{iso})
can be realized in several ways, depending on which matter 
component is initially perturbed, or, in other words, on 
which $\delta_{x}$ is initially different from zero. 
In the present case a further degree of freedom arises 
from the presence of the scalar field, and we will analyze separately 
two main situations: in the first case only one matter component 
(CDM or baryons) is initially perturbed; in the second case 
the initially perturbed component is only the scalar field.

\subsection{Isocurvature conditions from matter perturbations}

Let us consider first the case in which an initial density perturbation,
with amplitude $\delta_{c0}$, resides only on the CDM component.
By integrating (\ref{deltacdot}), one finds
\begin{equation}
\label{cdmtime}
\delta_{c}=\delta_{c0}-{1\over 2}h\ .
\end{equation}
By hypothesis, this is the only initially perturbed quantity.
All the others must  to be set to zero at $\tau =0$. 
Let us search the early time behaviour of the perturbations.
Since all the modes are outside the horizon at early times,
we first neglect all the terms proportional to $k$ in the Einstein
equations; then we expand all the quantities
in powers of $\epsilon$ defined in (\ref{epsilon}) and we calculate
the leading orders. In making this, we are assuming that
all the perturbation quantities admit a Taylor expansion in
$\tau =0$ of course. By making use of the above criteria and
of equations (\ref{aepsilon},\ref{Hepsilon}), the Einstein
equation (\ref{t00}) becomes 
$$
\left(1+{1\over 4}{\Omega_{m}\over\Omega_{r}}\epsilon +
O(\epsilon^{2}) \right)
\left(1+{1\over 2}{\Omega_{m}\over\Omega_{r}}\epsilon +
O(\epsilon^{2}) \right)\epsilon \dot{h}=
$$
\begin{equation}
\label{einsteindeveloped}
= {8\pi G\over {\cal C}^{2}}\left(\Omega_{\gamma}\delta_{\gamma}+
\Omega_{\nu}\delta_{\nu}+\Omega_{c}\delta_{c}\epsilon +
\Omega_{b}\delta_{b}\epsilon +O(\epsilon^{2})\right)\ , 
\end{equation}
and it is immediate to gain the early time behaviour of $h$:
\begin{equation}
\label{hearly}
h=\delta_{c0}{\Omega_{c}\over\Omega_{r}}\epsilon
-{3\over 8}\delta_{c0}{\Omega_{c}\Omega_{m}\over\Omega_{r}^{2}}
\epsilon^{2}+O(\epsilon^{3})\ .
\end{equation}
From the  arguments exposed at the beginning of this 
section, up to the order $\epsilon^{2}$ we have also
\begin{equation}
\label{etaearly}
\eta =-{1\over 6}h\ .
\end{equation}
Let us come now to the fluid perturbation quantities. 
As it is evident from the fluid equations, the $\theta$ and
$\sigma$ quantities are of higher order in $k\tau$ with respect
to the purely metric perturbations $h$ and $\eta$. Therefore,
their early time behaviour can be written as follows:
\begin{equation}
\label{fluidearly}
\delta_{\gamma}=\delta_{\nu}={4 \over 3}\delta_{b}=-{2\over 3}h\ ,
\end{equation}
\begin{equation}
\label{thetaearly}
\theta_{\gamma}=\theta_{\nu}=\theta_{b}=-{1\over 12}
\delta_{c0}{\Omega_{c}\over\Omega_{r}{\cal C}}k^{2}\epsilon^{2}+
O(\epsilon^{3})\ ,\ \theta_{c}=0\ ,\ \ \sigma_{\nu}=O(\epsilon^{3})\  .
\end{equation}
The behaviour of $h+6\eta$ is interesting even if of high order in 
$\tau$ since it is directly related to the gauge invariant 
curvature by (\ref{curv},\ref{psisync}) 
and it can be obtained by solving equation 
(\ref{tij}): 
\begin{equation}
\label{hp6eta}
h+6\eta ={{\cal I}_{1}\over 3}\tau^{3}\ ,
\end{equation}
where we have defined 
\begin{equation}
\label{i1}
{\cal I}_{1}={4\delta_{c0}\Omega_{c}(\Omega_{\nu}-5\Omega_{r})
{\cal C}k^{2}\over 36\Omega_{r}+24\Omega_{\nu}}\ .
\end{equation}
Note that $h+6\eta\propto\tau^{3}$, according to the isocurvature nature 
of the present case, as we exposed in the beginning of this section. 
Also, (\ref{hp6eta}) can be used to find the behaviour of $\sigma_{\nu}$, 
by using again (\ref{tij}). 

It remains to find the early time behaviour of the scalar
field perturbation $\delta\phi$. This can be done 
 by expanding $\delta\phi$ in powers of $\epsilon$ and looking at
the perturbed Klein Gordon equation once the terms proportional
to $k^{2}$ have been neglected. The inhomogeneous term is
$-{1 \over 2} \dot{\phi}\dot{h}$; $\dot{h}$ is of the order zero 
from (\ref{hearly}), and $\dot{\phi}=a\cdot\phi_{t}$
is at least of the order $\epsilon$; thereby, to the lowest order in
$\epsilon$, equation (\ref{KGp}) is satisfied by 
\begin{equation}
\label{deltaphiearly}
\delta\phi = -{1 \over 24} {\phi}_{t 0} {\delta}_{c0}
{\Omega_c \over \Omega_r }{ \epsilon^3 \over \cal{C} }
\ \ , \ \
\ \delta\dot{\phi}= -{1 \over 8} {\phi}_{t 0} {\delta}_{c0}
{\Omega_c \over \Omega_r }{ \epsilon^2 }.
\end{equation}
This completes the early time behaviour of all the perturbation
quantities in this case of isocurvature initial conditions.
All these relations can be easily generalized to the case in which
the initial perturbed matter component is the baryonic one.
In the next subsection we study the other interesting case:
the initial perturbed fluid component is the scalar field itself. 

\subsection{Isocurvature conditions from scalar field perturbations}

Let us suppose that at the initial time $a \rightarrow 0$ the only
non-zero perturbed quantity is the scalar field, in a manner such that 
the total  gauge-invariant energy density
contrast is zero, all the others perturbations being zero; 
this means that at least one of the two quantities 
$\delta_{\phi 0},\theta_{\phi 0}$ must be different from zero initially; 
from (\ref{deltaphithetaphi}), 
the corresponding expressions for $\delta\phi_{0}$ and 
$\delta\phi_{t0}$ are 
\begin{equation}
\delta\phi_{0}={1\over V'(\phi_{0})}
\left[{1\over 2}\phi_{t0}^{2}
\left(\delta_{\phi 0}-2{\theta_{\phi 0}\over k}\right)+
V(\phi_{0})\delta_{\phi 0}\right]\ ,\ \delta\phi_{t0}=
{\phi_{t0}\theta_{\phi 0}\over k}\ .
\end{equation}
In order to have isocurvature, for the other quantities we impose 
again the initial conditions (\ref{iso}). 
The relevant difference with respect to  the previous
situation lies in the  slower rise of the metric and fluid
perturbations starting from their initial zero values: they will grow
according to
(\ref{t00})-(\ref{tij}) and (\ref{system1}-\ref{system3}),  the whole
perturbation-growth-machinery being initially driven only by the
 $O(\epsilon^4)$ contribution of the scalar field through the
perturbed Einstein equations, while the perturbed Klein Gordon equation starts
 its dynamics from the conditions $\delta \phi_0\ne 0\ ,\ \delta
\dot{\phi}=0$ and generates the inhomogeneous term
driving the evolution of $h$.
From equation (\ref{t00}), together with (\ref{system1}-\ref{system3}), 
it is easy to
find the early-time behaviour of the metric perturbation $h$:
\begin{equation}
h={3 \over 4 } \left( { \delta_{\phi 0} \ \rho_{\phi}    \over
\rho_{c} \Omega_r          } \right) {\cal{C}}^4 \tau^4 +
O(\tau^5) \ .
\end{equation}
Using the method applied  in the previous sections, one finds  the
leading-order behaviours:  
\begin{equation}
\delta_{\gamma}= \delta_{\nu} = {4 \over 3 } \delta_c={4 \over 3 } \delta_b
= -{2 \over 3 }h \propto \tau^4
\end{equation}
\begin{equation}
\theta_{\nu} , \theta_{\gamma}, \theta_b \ \propto \tau^5 \ , \ \ \ 
\sigma_\nu \propto \tau^6\ ,
\end{equation} 
and, from (\ref{tij}) it can be seen that 
\begin{equation}
\label{hplus6eta2}
h+6\eta ={{\cal I}_{2}\over 6}\tau^{6}\ ,
\end{equation}
where 
\begin{equation}
\label{i2}
{\cal I}_{2}={k^{2}{\cal C}^{4}\over 170\Omega_{r}+8\Omega_{\nu}}
\left({\delta_{\phi 0}\rho_{\phi}\over\rho_{c}\Omega_{r}}\right)
\left(-{2\over 10}\Omega_{\nu}-{125\over 10}\Omega_{r}\right)\ .
\end{equation}
From the above formulas we see that the perturbations 
regarding the metric and the ordinary fluid components rise 
very slowly; indeed we found a substantial failure of this 
model in providing a significant amount of perturbations.
For this reason we will not consider this case in the 
numerical integrations of section VII. 

It is interesting to find 
the behaviour of scalar field perturbation at early times, 
that moves it away from its initial value $\delta\phi_{0}$; this 
contains corrections in $(k\tau^{2})$ together with a term 
proportional to $\tau^{4}$, as can be verified 
by integration of (\ref{KGp}): 
\begin{equation}
\label{phiisob}
\delta\phi =\delta\phi_{0}+\delta\phi^{(2)}\tau^{2}+
\delta\phi^{(3)}\tau^{3}+\delta\phi^{(4)}\tau^{4}+
\delta\phi^{(5)}\tau^{5}+\delta\phi^{(6)}\tau^{6}+O(\tau^{7})\ ,
\end{equation}
where the expansion coefficients are given by
$$
\delta\phi^{(2)}=-{1 \over  6} \delta\phi_{0} \ k^2\ \ ,\ \ 
\delta\phi^{(3)}={1\over 72}{\Omega_{m}\over\Omega_{r}}{\cal C}k^{2}
\delta\phi_{0}\ ,
$$
$$
\delta\phi^{(4)}={1\over 20}\delta\phi_{0}
\left({k^4 \over 6}- {\cal C}^{2}V''\right)+
{1\over 80}\left({\Omega_{m}\over\Omega_{r}}\right)^{2}
{\cal C}^{2}\delta\phi^{(2)}-
{3\over 40}{\Omega_{m}\over\Omega_{r}}{\cal C}\delta\phi^{(3)}\ ,
$$
$$
\delta\phi^{(5)}=-{1\over 15}{\Omega_{m}\over\Omega_{r}}{\cal C}
\delta\phi^{(4)}+\left[{1\over 80}
\left({\Omega_{m}\over\Omega_{r}}\right)^{2}{\cal C}^{2}-k^{2}
\right]\delta\phi^{(3)}\ ,
$$
\begin{equation}
\label{coeffphiisob}
\delta\phi^{(6)}=-{5\over 84}{\Omega_{m}\over\Omega_{r}}{\cal C}
\delta\phi^{(5)}+\left[{1\over 84}
\left({\Omega_{m}\over\Omega_{r}}\right)^{2}{\cal C}^{2}-k^{2}
\right]\delta\phi^{(4)}-{\cal C}^{2}V''\delta\phi^{(2)}-
{3\over 2}\left({\delta_{\phi 0}\rho_{\phi}\over
\rho_{c}\Omega_r} \right) {\cal{C}}^{5}\phi_{t0}\ .
\end{equation}
We considered the expansion up to the sixth order in $\tau$ 
because, as we will see in the next section, going to the 
Newtonian gauge changes the last coefficient. 

In the next section we extend the results of section IV and V to 
the conformal Newtonian gauge. 

\section{Results in the Conformal Newtonian gauge}

As it is well known, the synchronous gauge is a 
coordinate system corresponding to observers at 
 rest with respect to the collisionless  matter component. 
These "Lagrangian
coordinates" are defined by the rest frame of a set of preferred
observers. More physical intuition can be achieved 
in the conformal Newtonian gauge, where the
metric tensor is diagonal. Inside the horizon, the perturbation
equations reduce to the standard non-relativistic Newtonian equations.
In this section  we write   the results of section IV and V 
in the  Newtonian gauge. 

The connection between two gauges is realized in general by performing 
a coordinate transformation relating the two frames. The link 
between the perturbations in the two gauges is expressed 
in the same coordinate point instead of the same spacetime point; 
this is why in most cases it is interesting to know the difference 
of the fluctuations in the two gauges \cite{KS}. 

First we write down the relations between the genuine metric 
perturbed quantities. In the Newtonian gauge the perturbation to 
$g_{00}$ exists and it is represented by the potential $\Psi$; 
the trace of $g_{ij}$ is instead perturbed by $\Phi$. 
Their relation with $h$ and $\eta$ are 
\begin{equation}
\label{psisn}
\Psi = {1\over 2k^{2}}\left[\ddot{h}+6\ddot{\eta}+{\dot{a}\over a}
\left(\dot{h}+6\dot{\eta}\right)\right]\ ,
\end{equation}
\begin{equation}
\label{phisn}
\Phi = {1\over 2k^{2}}{\dot{a}\over a}
\left(\dot{h}+6\dot{\eta}\right)-\eta\ .
\end{equation}
They can be easily expressed for $\epsilon\ll 1$,$k\tau\ll 1$ 
by substituting directly the expressions for $h$ and $\eta$ 
contained in sections IV and V. 

Now we concentrate on the transformations regarding fluids and scalar 
field. They are the contained in the stress energy tensor, that 
transforms as 
\begin{equation}
\label{tmunugauge}
T^{\mu\nu}(\tilde{x})={\partial\tilde{x}^{\mu}\over\partial x^{\rho}}
{\partial\tilde{x}^{\nu}\over\partial x^{\sigma}}T^{\rho\sigma}(x)\ .
\end{equation}
Using this and taking care to compare the perturbations in the same 
coordinate point, the relations between the quantities in the two 
gauges (synchronous and Newtonian labeled as $s$ and $N$ respectively) 
for each fluid are:
\begin{eqnarray}
\label{deltagauge}
\delta_{s}&=&\delta_{N}-{\cal T}{\dot{\rho}\over\rho}\ ,\\
\label{thetagauge}
\theta_{s}&=&\theta_{N}-k^{2}{\cal T}\ ,\\
\label{pgauge}
p_{s}&=&p_{N}-p{\cal T}\ ,\\
\label{sigmagauge}
\sigma_{s}&=&\sigma_{N}\ ,
\end{eqnarray}
where 
\begin{equation}
\label{}
{\cal T}={\dot{h}+6\dot{\eta}\over 2k^{2}}
\end{equation}
is the lapse between the synchronous and the Newtonian time coordinate. 
Regarding the scalar field, we compute the Newtonian gauge expression  
of the amplitude fluctuation $\delta\phi$ by using the transformation 
\begin{equation}
\label{phigauge}
\delta\phi_{s}=\delta\phi_{N}-\dot{\phi}{\cal T}\ .
\end{equation}
In the following subsections we write 
the behaviour of the fluid quantities 
in the $\epsilon\ll 1$,$k\tau\ll 1$ regime and in the 
Newtonian gauge, dropping the $N$ subscript. 

\subsection{Adiabaticity}

The leading orders for matter and radiation perturbations are 
\begin{equation}
\label{adidelta}
\delta_{\gamma}=-{40{\cal N}\over 15+4R_{\nu}}=\delta_{\nu}=
{4\over 3}\delta_{b}={4\over 3}\delta_{c}\ ,
\end{equation}
\begin{equation}
\label{aditheta}
\theta_{\gamma}=\theta_{\nu}=\theta_{c}=\theta_{b}=
{10{\cal N}\over 15+4R_{\nu}}k^{2}\tau \ ,
\end{equation}
\begin{equation}
\label{adisigma}
\sigma_{\nu}={4{\cal N}\over 45+12R_{\nu}}k^{2}\tau^{2}\ .
\end{equation}
The scalar field perturbation amplitude is 
\begin{equation}
\label{adiphi}
\delta{\phi}={\cal N}\tau^{2}\phi_{t0}{10\over 15+4R_{\nu}}\ .
\end{equation}
Note that in this case the scalar field perturbations grow 
in time faster ($\propto\tau^{2}$) than in the synchronous 
gauge ($\propto\tau^{4}$); as we point out below, 
this is not a feature of the isocurvature initial conditions. 

\subsection{Isocurvature from matter}

Matter and radiation behave as 
\begin{equation}
\label{isomdelta}
\delta_{\gamma}=\delta_{\nu}={4\over 3}\delta_{b}=
-{2\over 3}\delta_{c0}{\Omega_{c}\over\Omega_{r}}{\cal N}\tau -
{2{\cal I}_{1}\over k^{2}}\tau\ .
\end{equation}
\begin{equation}
\label{isomdeltac}
\delta_{c}=\delta_{c0}-{1\over 2}\delta_{c0}{\Omega_{c}\over\Omega_{r}}
{\cal C}\tau -3{{\cal I}_{1}\tau\over 2k^{2}}\ ,
\end{equation}
\begin{equation}
\label{isomtheta}
\theta_{\gamma}=\theta_{\nu}=\theta_{b}=
-{1\over 12}\delta_{c0}{\Omega_{c}\over\Omega_{r}}k^{2}
{\cal C}\tau^{2}+{{\cal I}_{1}\tau^{2}\over 2}
\ ,\ \theta_{c}={{\cal I}_{1}\tau^{2}\over 2}\ .
\end{equation}
\begin{equation}
\label{isomsigma}
\sigma_{\nu}=O(\epsilon^{3})\ .
\end{equation}
The scalar field amplitude is given by 
\begin{equation}
\label{isomphi}
\delta\phi =-{1\over 24}\phi_{t0}\delta_{c0}
{\Omega_{c}\over\Omega_{r}}{\cal C}^{2}\tau^{3}+
{{\cal I}_{1}{\cal C}\tau^{3}\over 2k^{2}}\phi_{t0}\ ,
\end{equation}
and shows that the gauge change does not touch the 
order of the leading power in $\tau$, 
although it modifies its numerical coefficient.  

\subsection{Isocurvature from scalar field}

Matter and radiation behave as 
\begin{equation}
\label{isosf}
\delta_{\gamma}=\delta_{\nu}={4\over 3}\delta_{b}=-{1\over 2}
\left({\delta_{\phi 0}\rho_{\phi}\over\rho_{c}\Omega_{r}}\right)
{\cal C}^{4}\tau^{4}-2{{\cal I}_{2}\over k^{2}}\tau^{4}\ ,
\end{equation}
\begin{equation}
\label{isosfdeltacthetac}
\delta_{c}=-{3\over 8}
\left({\delta_{\phi 0}\rho_{\phi}\over\rho_{c}\Omega_{r}}\right)
{\cal C}^{4}\tau^{4}-{3{\cal I}_{2}\tau^{4}\over 2k^{2}}
\ ,\ \theta_{c}={{\cal I}_{2}\tau^{5}\over 2}\ ,
\end{equation}
\begin{equation}
\label{isosfthetasigma}
\theta_{\gamma}=\theta_{\nu}=\theta_{b}\propto\tau^{5}
\ ,\ \sigma_{\nu}\propto\tau^{6}\ .
\end{equation}
The scalar field amplitude is given by 
\begin{equation}
\label{phiisobn}
\delta\phi =\delta\phi_{0}+\delta\phi^{(2)}\tau^{2}+
\delta\phi^{(3)}\tau^{3}+\delta\phi^{(4)}\tau^{4}+
\delta\phi^{(5)}\tau^{5}+
\left(\delta\phi^{(6)}+
{{\cal I}_{2}\over 2k^{2}}\phi_{t0}\right)\tau^{6}+O(\tau^{7})\ ;
\end{equation}
therefore, in this isocurvature case, the behaviour of 
$\delta\phi$ is affected by the gauge change only at high orders 
in $\tau$, leaving the leading terms unperturbed. 

In the next section we will numerically solve the linear
cosmological perturbation equations with the initial conditions sets
developed in the IV and V sections. 

\section{Numerical integrations and discussion}

We performed the numerical integration applying our considerations 
to a scalar field model based on ultra-light pseudo-Nambu-Goldstone 
bosons; the potential associated to this field has the 
form \cite{nambu}:
\begin{equation}
V(\phi)= M^4 [cos(\phi /f)+1]
\end{equation}
Our working point corresponds to the parameter choice $f=1.885 \times
10^{18}$ GeV and $M=10^{-3}$ eV; assuming an initial kinetic energy
equal to the potential one, the starting values of the scalar field
and its initial time derivative are obtained by requiring that the
present contribution is $\Omega_{\phi}=0.6$, fixing $H_0=70$ km/sec/Mpc 
and $\Omega_b=0.05$. Furthermore, we have taken the primordial power
spectrum to be exactly scale-invariant. 

Even though the main cosmological consequences of this kind of scalar
fields have been analyzed by many authors (see \cite{CDF,VL}), 
here we use the formulas developed in the previous sections 
to accurately compare the pure adiabatic and pure isocurvature regimes. 
Also we give particular emphasis on the behaviour of
entropy and curvature perturbations, again comparing their 
evolution starting from isocurvature (from CDM) and adiabatic 
initial conditions; each case is compared with a pure CDM model 
with the same background parameters. 

First, let us consider the power spectra of the microwave 
background anisotropies, both temperature and 
polarization. They are expressed by the expansion 
coefficients of the two point correlation function into 
Legendre polynomials (see e.g. \cite{PAD}) and admit the 
following expression in terms of the quantities defined 
in the previous sections: 
\begin{equation}
\label{spectra}
C_{l}^{T}=4\pi 
\int {dk\over k}|\Delta_{Tl}(k,\tau_{0})|^{2}\ \ ,
C_{l}^{P}=4\pi
\int {dk\over k}|\Delta_{Pl}(k,\tau_{0})|^{2}\ \ .
\end{equation}
The adiabatic case is shown in 
figure 1. The presence of the scalar field (solid line) 
produces an increase of the power of the acoustic oscillations 
with respect to the CDM model (dashed line); this is due to the fact 
that the universe is not completely matter dominated at decoupling: 
thus at this time the perturbations are growing faster than in the 
CDM models (we recall that density perturbations in adiabatic models 
grow as $a^{2}$ and $a$ respectively in the radiation in the matter eras) 
and this produces an early integrated Sachs-Wolfe effect (ISW), 
found first in \cite{CDF}. Also, the position of the first peak 
is slightly shifted toward smaller angular scales due 
to the increase of the distance of the last scattering  
surface (projection effect). Note how these features 
regard both the polarization and temperature peaks. 
Finally, the temperature spectra shows that the ISW is active 
on the smallest multipoles due to the dynamics of the 
scalar field in the present case; this is a distinctive 
feature with respect to the cosmological constant \cite{CDF}. 

Figure 2 shows the spectrum from isocurvature perturbations. 
While the projection effect is the same as in the adiabatic case, 
now the situation regarding the amplitude of the 
acoustic oscillations is inverted: the peaks are lower than 
the ordinary models, both for polarization and temperature. 
This is simply due to the reduction of the matter/radiation ratio as 
we include $\phi$ by keeping $\Omega_{total}=1$; in fact, 
the scalar field has {\it no} intrinsic dynamical effect 
at last scattering since matter 
and radiation components were largely dominant: 
it is well known the opposite behavior of the anisotropies 
in adiabatic and isocurvature models as one varies 
$\Omega_{m}h^{2}$ (see e.g. \cite{PAD}). 
To better see this point, 
we plot in figure 3 and 4 the power spectra for 
models having fixed $\Omega_{b}h^{2}$ and $\Omega_{c}h^{2}$ 
but varying $\Omega_{m}=\Omega_{b}+\Omega_{c}$ and 
$h$ by mean of different $\Omega_{\phi}$. 
Thus we expect the same amount of 
perturbations in the CMB except for the effects that are 
genuinely linked to the scalar field, as the projection effect 
and the ISW on the smallest multipoles. This is precisely 
what happens for the spectra in figure 3 and 4. The 
dashed lines represent again the curves for 
$\Omega_{\phi}=.6$ as in figures 1 and 2; the solid and 
thin lines represents $\Omega_{\phi}=.5$ and $.7$ respectively. 
Again, the spectra show remarkably the same features 
for polarization and temperature, even if it should be 
noted how the former, arising from acoustic 
oscillations occurring just {\it at} decoupling, 
is not influenced by the ISW effect, since it comes 
from the line of sight integration. All 
this shows how the perturbation power at decoupling 
is not touched by the subdominant scalar field; the opposite 
behaviours in the adiabatic and isocurvature cases is 
explained by the decrease of matter in favor of scalar field. 

More insight on the perturbation behaviours may be 
obtained by looking at the time evolution of 
some significant quantities; we look at one single 
scale, or wavenumber, roughly entering in the horizon 
between matter and radiation equality and decoupling: 
\begin{equation}
\label{k}
k=8\times 10^{-2} {\rm Mpc}^{-1}\ .
\end{equation}
Let us begin with the gravitational potential $\Psi$, 
in figure 5; the oscillatory dynamics is associated, 
both in the isocurvature and adiabatic case, to 
the horizon crossing of the scale 
examined. The amplitude of the oscillations in the scalar 
field models are higher for the adiabatic case and 
lower for the isocurvature one when compared with the 
corresponding CDM models. These oscillations are the 
source of the CMB anisotropies on sub-horizon angular scales 
($l\ge 200$) through the acoustic 
driving effect and the early ISW effect 
\cite{HSS}, and therefore follow the different behaviour in the 
two cases. 

We concentrate now on two particularly significant 
quantities regarding both adiabatic and isocurvature 
regimes, the total entropy perturbation, defined below, 
and the curvature $\zeta$ defined in (\ref{curv}); 
we recall that these quantities are gauge invariant. 
The amplitude of the total entropy perturbation is given by
\begin{eqnarray}
p \Gamma= p {\Gamma}_{int} + p{\Gamma}_{rel} &=&
\sum _a (\delta p_a -c_a^2 \delta \rho _a)+ \sum _a (c_{a}^2-c_s^2) \delta
\rho _a = \nonumber\\
\label{Gamma}
 &=& \sum _a ( \delta p_a  - c_s^2 \delta \rho _a )\ ,
\end{eqnarray}
where $p$ is the total pressure and the sound speed must take into
account the scalar field contribution 
$c_{\phi}^2=1-2 V' \dot{\phi}/\dot{\rho}_{\phi}$ in the summation: 
\begin{equation}
\label{cs2}
c_s^2={ \sum_a h_a c_a^2  \over h}\ ,
\end{equation}
where 
\begin{equation}
h_a= \rho_a+p_a \ \ \ \ , \ \ \ \ h=\sum_a h_a\ .
\end{equation}
While the $\Gamma_{int}$ term comes from the intrinsic entropy
perturbation of each component,  and it is ultimately sourced only by the
scalar field  component due to the spatial and temporal variations of the
scalar field equation of state, the $\Gamma_{rel}$ term arises from the 
different dynamical behaviours of
the components, and it is related to the $S_{ab}$ quantities defined
in eq. (\ref{sab}) by the relation \cite{KS}:
\begin{equation}
p\Gamma_{rel}={1 \over 2} \sum_{a,b} 
{h_a h_b \over h} (c_a^2 - c_b^2)S_{ab}
\end{equation}
In fig.6 and 7 we plot $\Gamma$ vs $a$ (solid line and arbitrary 
units) using isocurvature CDM and adiabatic initial conditions respectively.
We note that in both cases $\Gamma \rightarrow 0$ as $a \rightarrow 0$: 
in the last case, this is an obvious consequence of the definition of what
adiabatic conditions are; on the other hand, taking isocurvature CDM
initial conditions, we started with non-zero values of the $S_{ab}$ relative
to the CDM component and the other non-compensating components, but again 
$\Gamma \rightarrow 0$ as $a \rightarrow 0$. This is because 
at early times the cosmic fluid is radiation dominated, 
so that $p\propto 1/a^{4}$ in (\ref{Gamma}); this kills 
$\Gamma$ for $a\rightarrow 0$, since no radiation perturbations 
are present initially. Instead, the initial value of the first time 
derivative of $\Gamma$ is different from zero only selecting 
isocurvature initial conditions, since 
it takes contributions directly from the $S_{ab}$ terms \cite{KS}. 
The behaviours of entropy perturbations in both cases have been compared
with those in the standard Einstein-De Sitter model $\Omega_m=1$ (dotted
line), with the same choice of $\Omega_b $ and $H_0$. The entropy 
perturbations remain nearly constant before horizon crossing; at 
this time the perturbation starts its oscillations that are damped 
in amplitude when the scale is well below the effective horizon. 
As an expected feature, note that in the scalar-field model 
the peaks of the oscillations are shifted closer to
the present when compared to the Einstein-De Sitter case, as the epoch of
matter-radiation equality. 

In fig.8 and 9 we plot the evolution of the gauge-invariant 
curvature perturbation $\zeta$ for isocurvature CDM and adiabatic 
initial conditions, respectively. At $a\ll 1$, this quantity is 
zero in the isocurvature case and non-zero in the adiabatic one, 
being an explicit indicator of the nature of the perturbations. 
Again the significant dynamics occurs in correspondence of the 
horizon crossing, and the latter occurs slightly later 
then in the CDM model due to the presence of the scalar field. 

Finally, note how in all the isocurvature cases (figures 6 and 8) 
the amplitude of the oscillations are lower than in the 
corresponding CDM models; as we mentioned above, 
this is due to the reduction of the matter component in favor of 
the scalar field energy density. Most important, 
these graphs show that this is the only 
possible cause of this effect, since at the times of oscillations 
for the scale examined, roughly between equivalence and decoupling, 
the scalar field is very subdominant with respect to the 
other components. 

The hypothesis of a cosmic vacuum energy stored 
in the potential of a scalar field enlarges naturally the 
possibility to gain insight into high energy physics 
from the traces left in the cosmic radiation and 
in the matter distribution. 
Due to the upcoming CMB experiments \cite{MP}, 
it will be interesting to further study the 
cosmological imprints of these models, in the context 
of different theories attempting to describe the hidden 
sector of high energy physics. 

\acknowledgments

We are grateful to Luigi Danese and 
Sabino Matarrese for their hints and encouragement.
We warmly thank Marco Bruni, Scott Dodelson, Andrew Liddle, 
Ed Copeland and Elena Pierpaoli for quick and accurate comments on 
our questions. Also, we are grateful to the anonymous referee for 
the precious comments and suggestions. C.B. wishes to thank 
the SISSA/ISAS institute for the kind hospitality.

\begin{figure}
\hskip .3in
\psfig{figure=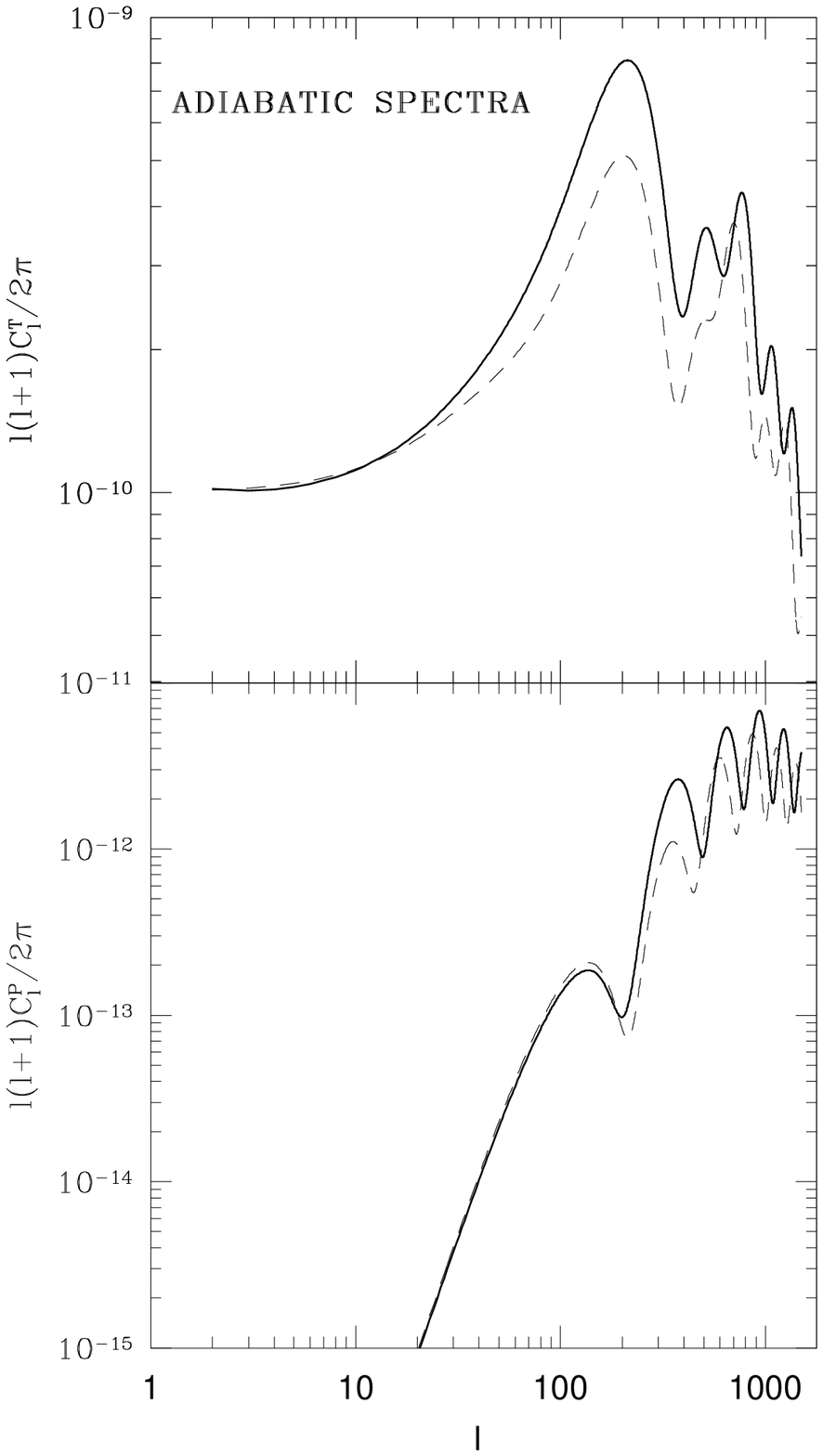,height=5in}
\caption{Power spectra of the CMB anisotropies from adiabatic 
initial conditions. The background parameters are $\Omega_{b}=.05$, 
$h=.7$, three massless neutrino families and 
$\Omega_{\phi}=.6,\Omega_{c}=.35$ (solid line), 
$\Omega_{\phi}=0,\Omega_{c}=.95$ (dashed dotted line). 
Note the increase of the acoustic peaks power 
in the scalar field model.}
\end{figure} 

\begin{figure}
\psfig{figure=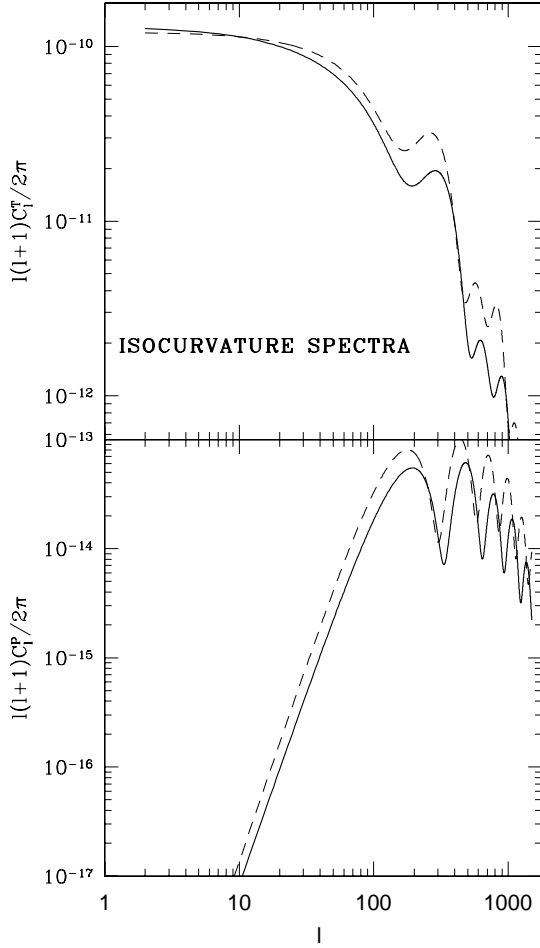,height=5in}
\caption{Power spectra of the CMB anisotropies from isocurvature 
initial conditions. The background parameters are $\Omega_{b}=.05$, 
$h=.7$, three massless neutrino families and 
$\Omega_{\phi}=.6,\Omega_{c}=.35$ (solid line), 
$\Omega_{\phi}=0,\Omega_{c}=.95$ (dashed dotted line). 
Note the decrease of the acoustic peaks power 
in the scalar field model, an opposite behaviour with respect 
to the adiabatic case.}
\end{figure} 

\begin{figure}
\psfig{figure=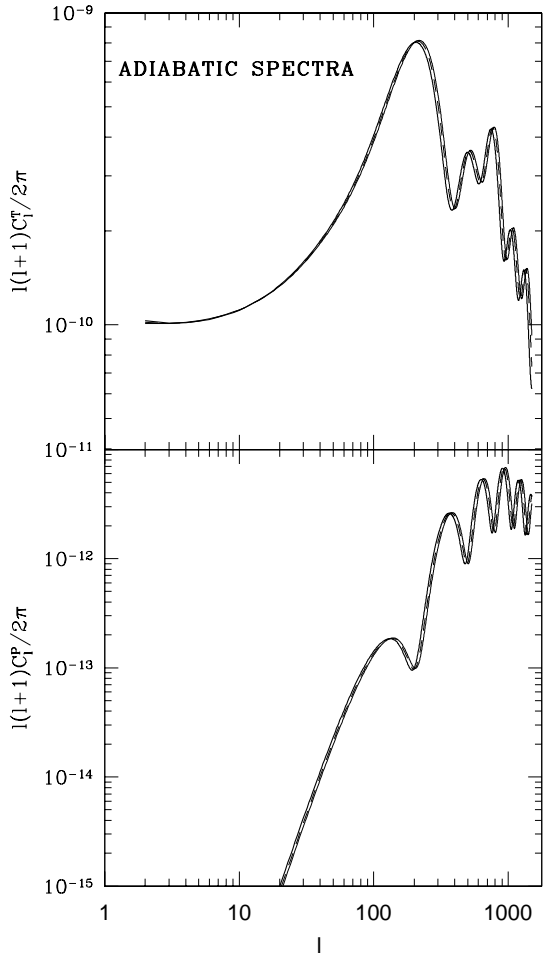,height=5in}
\caption{Power spectra of the CMB anisotropies from adiabatic 
initial conditions with different $\Omega_{\phi}$ 
and fixed $\Omega_{b}h^{2}=.0245$ and 
$\Omega_{c}h^{2}=.1715$ ($\Omega_{m}h^{2}=0.196$) 
as in figure 1 and 2. 
The background parameters are $\Omega_{m}=.4,h=.7$ 
(dashed line), $\Omega_{m}=.3,h=.81$ (thin line), and 
$\Omega_{m}=.5,h=.63$ (thick line). The amplitude 
of the peaks is the same, while they are 
slightly shifted because of the projection effect.}
\end{figure} 

\begin{figure}
\psfig{figure=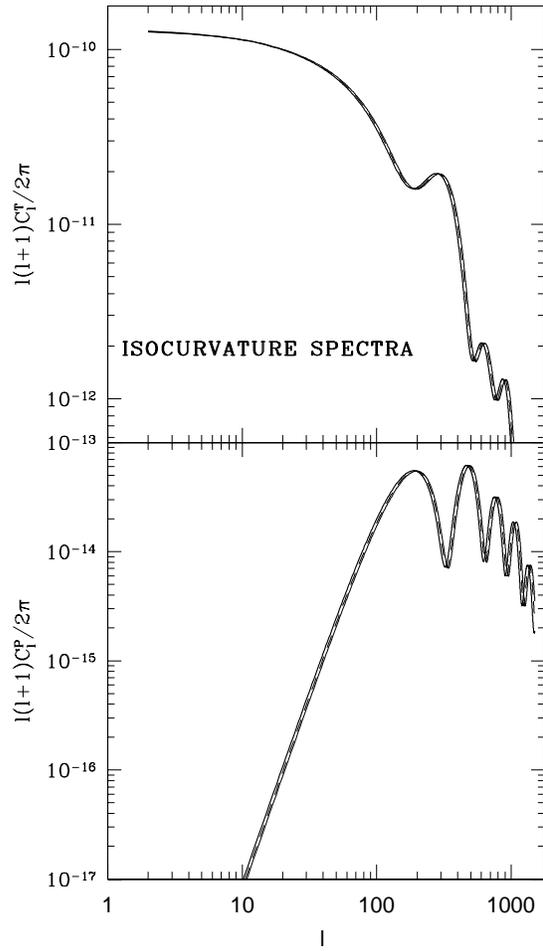,height=5in}
\caption{Power spectra of the CMB anisotropies from isocurvature 
initial conditions with different $\Omega_{\phi}$ 
and fixed $\Omega_{m}h^{2}=0.196$. The spectra 
show the same behaviours for varying $\Omega_{m}$ 
as in figure 3.}

\end{figure} 

\begin{figure}
\psfig{figure=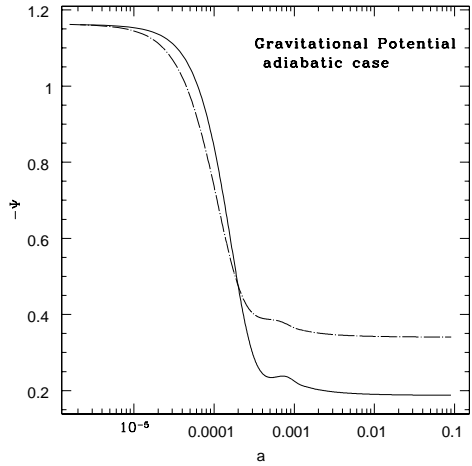,height=3.5in}
\psfig{figure=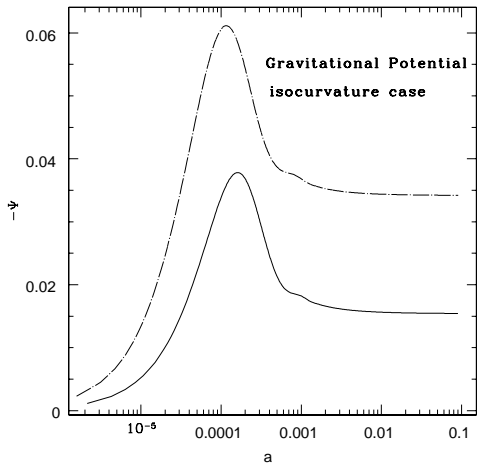,height=3.5in} 
\caption{Gravitational potential (in arbitrary units) 
at the comoving wavenumber $k=8\times 10^{-2}$ Mpc$^{-1}$ 
as a function of the time for adiabatic (top) and 
isocurvature (bottom) initial conditions. 
The background parameters are the same 
as in figures 1 and 2. The oscillatory dynamics is associated 
with the horizon crossing of the scale considered. 
In the adiabatic (isocurvature) 
scalar field models, the oscillation amplitudes are 
larger (lower) than in the corresponding CDM cases, 
according to the power spectra behaviours.}
\end{figure} 

\begin{figure}
\psfig{figure=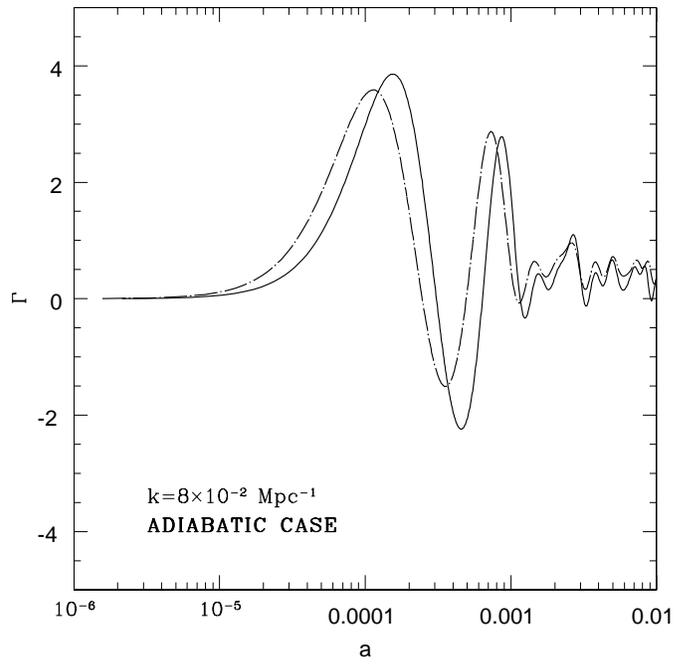,height=5in}
\caption{Gauge invariant entropy behaviour 
(in arbitrary units) as 
a function of the time in adiabatic models for scalar field 
(solid line) and pure CDM (dashed line) models. 
Note the shift of the horizon crossing (corresponding 
to the oscillations) toward late times due to the effective 
cosmological constant.}
\end{figure} 

\begin{figure}
\psfig{figure=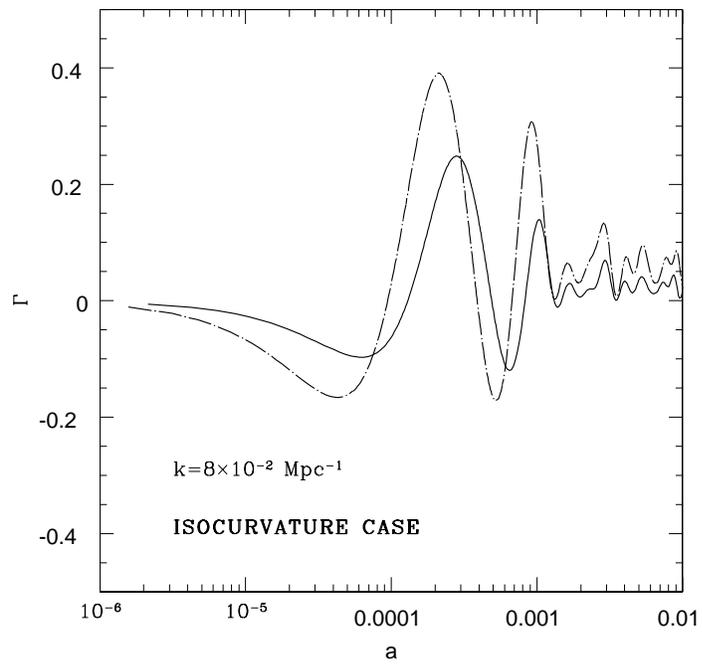,height=5in}
\caption{Gauge invariant entropy behaviour 
(in arbitrary units) as 
a function of the time in isocurvature models for scalar field 
(solid line) and pure CDM (dashed line) models. 
Note the decrease of the oscillation amplitudes in scalar field 
models, due to the lack of matter with respect to the pure 
CDM case.}
\end{figure}
 
\begin{figure}
\psfig{figure=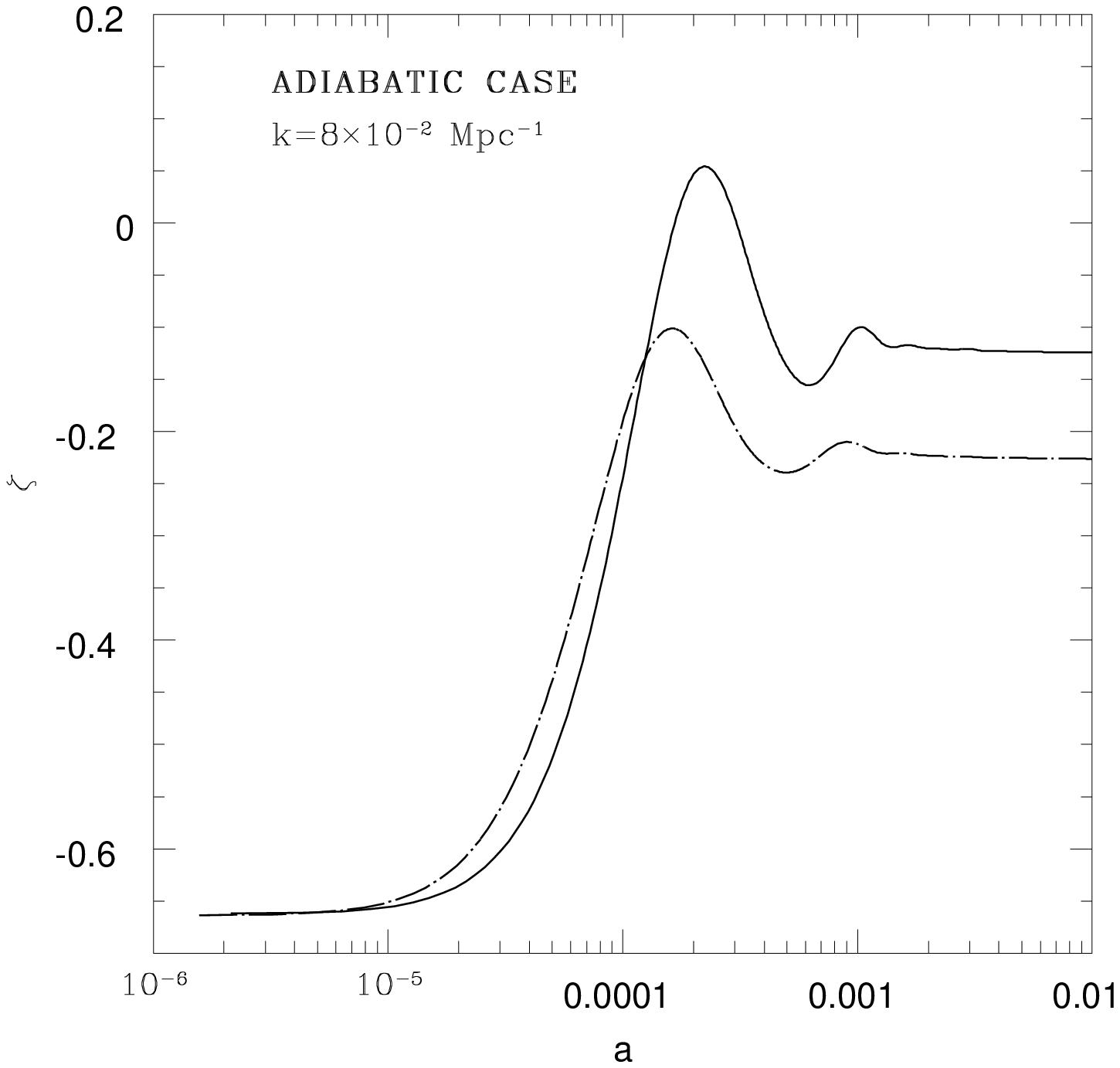,height=5in}
\caption{Gauge invariant curvature behaviour 
(in arbitrary units) as 
a function of the time in adiabatic models for scalar field 
(solid line) and pure CDM (dashed line) models. 
Note that the curvature is non-vanishing as $a\rightarrow 0$.}
\end{figure} 

\begin{figure}
\psfig{figure=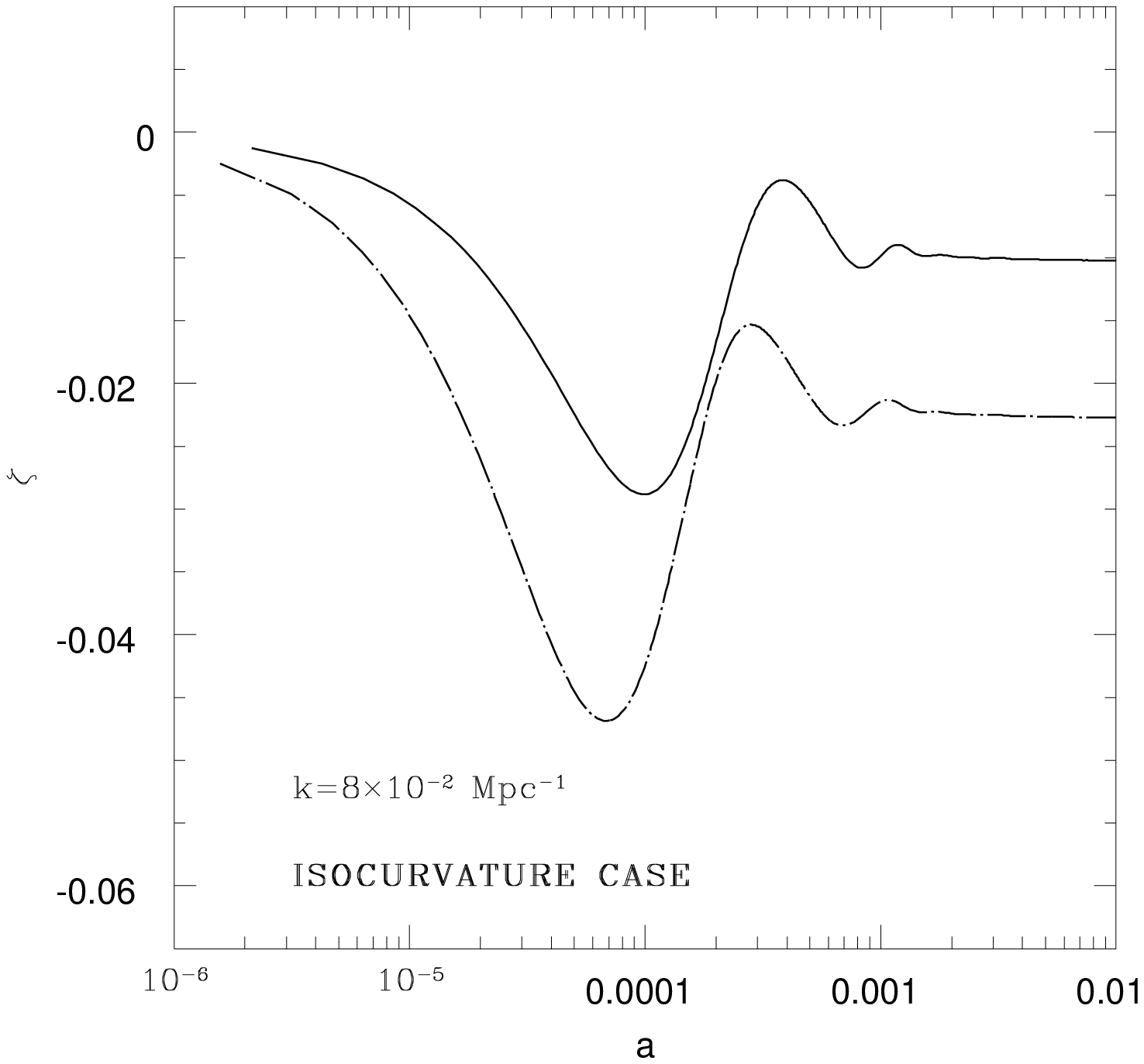,height=5in}
\caption{Gauge invariant curvature behaviour 
(in arbitrary units) as 
a function of the time in isocurvature models for scalar field 
(solid line) and pure CDM (dashed line) models. 
Note that the curvature is vanishing as $a\rightarrow 0$.}
\end{figure}

\today

\end{document}